\documentclass[aps,12pt,preprint]{revtex4} 
\usepackage{graphicx}
\usepackage{cancel}
\usepackage{amssymb}
\def\lsim{\mathrel{\raise.3ex\hbox{$<$\kern-.75em\lower1ex\hbox{$\sim$}}}}
\def\gsim{\mathrel{\raise.3ex\hbox{$>$\kern-.75em\lower1ex\hbox{$\sim$}}}}

\def\be{\begin{equation}}
\def\ee{\end{equation}}                         
\def\bea{\begin{eqnarray}}
\def\eea{\end{eqnarray}}   
 
\def\journal#1#2#3#4{{ #1} {\bf #2}, #3 (#4)}

\def\prd{Phys. Rev. D}

\begin{document}

\preprint{MADPH 08-1524}
\preprint{IPMU 08-0068}
\preprint{CAS-KITPC/ITP-072}

\title{GeV Majorana Neutrinos in Top-quark Decay at the LHC}

\author{Zongguo Si $^{a,d}$}
\email{zgsi@sdu.edu.cn}
\author{Kai Wang $^{b,c,d}$}
\email{kai.wang@ipmu.jp}
\affiliation{
$^a${School of Physics, Shandong University, Jinan, Shandong 250100, CHINA}\\
$^b${Institute for the Physics and Mathematics of the Universe, University of Tokyo, Kashiwa, Chiba 277-8568, JAPAN}\\
$^c${Department of Physics, University of Wisconsin, Madison, Wisconsin 53706, USA}\\
$^d${Kavli Institute for Theoretical Physics China, CAS, Beijing 100190, CHINA}
}

\begin{abstract}
We explore the $\Delta L=2$ same-sign dilepton signal 
from top-quark decay via a Majorana neutrino at the LHC in the top anti-top pair
production samples. The signature is same-sign dilepton plus multi-jets with no significant missing energy.
The most optimistic region lies where the Majorana neutrino mass is between $15-65$ GeV. For 300 fb$^{-1}$ integrated luminosity,
it is possible to probe $S_{ij}$, the effective mixing parameter,
to ${\mathcal O}(10^{-5})$. 
\end{abstract}

\maketitle

\section{Introduction}

Evidence for neutrino mass clearly indicates the need for new physics beyond 
standard model (SM)\cite{review}. The $10^{12}$ order hierarchy in 
$m_t/m_\nu$ and the large mixing in the neutrino sector also suggest a possible 
different mechanism for neutrino mass generation from the SM Yukawa interactions.
In addition, its electric neutrality allows for the 
possibility of neutrinos being Majorana fermions. Consequently, $\Delta L = 2$ lepton number violation (LNV)
will always occur in those theories \cite{Weinberg,TypeI,TypeII}.  
Taking an effective theory approach, Majorana neutrino mass generation can be categorized into
a SM gauge invariant non-renormalizable operator \cite{Weinberg}
$$
\lambda \ell\ell H H/\Lambda_{\cancel{L}},
$$
where the $\ell$ and $H$ are $SU(2)$ doublets, $\Lambda_{\cancel{L}}$ is the new physics scales at which 
lepton number violation occurs. The smallness of neutrino masses then suggests a large $\Lambda_{\cancel{L}}$. 
Various neutrino models have employed this so-called ``seesaw"
spirit \cite{TypeI,TypeII}. For instance, given $\lambda\sim{\mathcal O}(1)$, 
the LNV scale $\Lambda_{\cancel{L}}$ needs to be $M_{\text{GUT}}$ to 
obtain $m_\nu \sim 0.1~\text{eV}$. This can be realized in a Type-I seesaw model \cite{TypeI} where
a standard model singlet Majorana neutrino $N^c$ is introduced per generation and the interaction is as
$$
\ell N^c H + M_N N^c N^c.
$$ 

The Large Hadron Collider (LHC) at CERN will soon
provide a great opportunity for exploring physics 
at TeV scale. There were recently several proposals to test the neutrino mass generation mechanisms at the LHC 
where the new physics responsible for neutrino mass generation is of ${\mathcal O}(10-10^3~\text{GeV})$.
For instance, in some extended Type-I models, Majorana neutrino $N$ may be accessible at the LHC \cite{Han:2006ip,lhctypei}.
Following the same notation in \cite{Han:2006ip}, in the presence of three Majorana neutrino states, the neutrino gauge 
eigen state can be written as 
\begin{equation}
\nu_{iL} = \sum^3_{m=1} U_{im}\nu_{mL} + \sum^6_{m'=4} V_{im'} N^c_{m'L}~,
\label{mix}
\end{equation}
where $i=e,\mu,\tau$. 
Therefore, the interaction between charged lepton and Majorana neutrino
mass eigen states is as:    
\begin{equation}
{\mathcal L} = -\frac{g}{2\sqrt{2}}V_{ij}W^+_\mu l_i\gamma^\mu(1-\gamma_5)N^c_j+h.c.~.
\end{equation}
In the conventional Type-I seesaw model where $M_N$ is of order $10^{14}$ GeV, 
the mixing $V_{ij}$ are highly suppressed. However, in some extended Type-I models, 
this constraint can be released \cite{Han:2006ip}. Here, we adopt the philosophy in \cite{Han:2006ip} 
by taking a pure phenomenology approach without assuming any a-proiri relationship among
 the mass and mixing parameters.

This interaction will lead to direct production of Majorana neutrinos.
The signal consists of dijet plus same-sign dilepton associated with no significant
 $\cancel{E}_T$, 
$$
q\bar{q}'\rightarrow l^\pm N \rightarrow l^\pm l^\pm (W^\mp)^* \rightarrow l^\pm l^\pm jj. 
$$

Currently, the Majorana nature of neutrinos is being tested at neutrinoless double beta decay experiments($0\nu\beta\beta$) \cite{00nubeta} and 
it provides the strongest bound on $V_{eN}$ as \cite{Han:2006ip}
\begin{equation}
\sum_N \frac{|V_{eN}|^2}{M_N}< 5\times 10^{-8} \text{GeV}^{-1}.
\end{equation}
The CERN LEP experiment 
suggests $|V_{\mu N}|^2$, $|V_{\tau N}|^2$ $\lessapprox 10^{-4}-10^{-5}$ for $M_N\sim 5-80~\text{GeV}$\cite{L3,DELPHI-OPAL,Han:2006ip}. 
The D$\cancel{0}$ and CDF detectors at Tevatron have also performed a direct search the light Majorana neutrino \cite{tevatronzhang}. 

The LHC is a ``top factory" with a NLO production rate of about 800 pb and single top 
rate of about 400 pb. In this top rich environment, 
similar to $W^\pm \rightarrow l^\pm N \rightarrow l^\pm l^\pm (W^\mp)^*$, 
we explore top decay into $N^c$. The unique signal final state which consists of same-sign 
dilepton with no significant $\cancel{E}_T$ makes the discovery possible. In the second section, we will discuss the top decay into Majorana neutrino. Finally in the third section, we will study this specific decay mode in $t\bar{t}$ pair production at the LHC. 

\section{Top-quark decay to a Majorana neutrino}

\begin{figure}[!tb]
\includegraphics[scale=1,width=5cm]{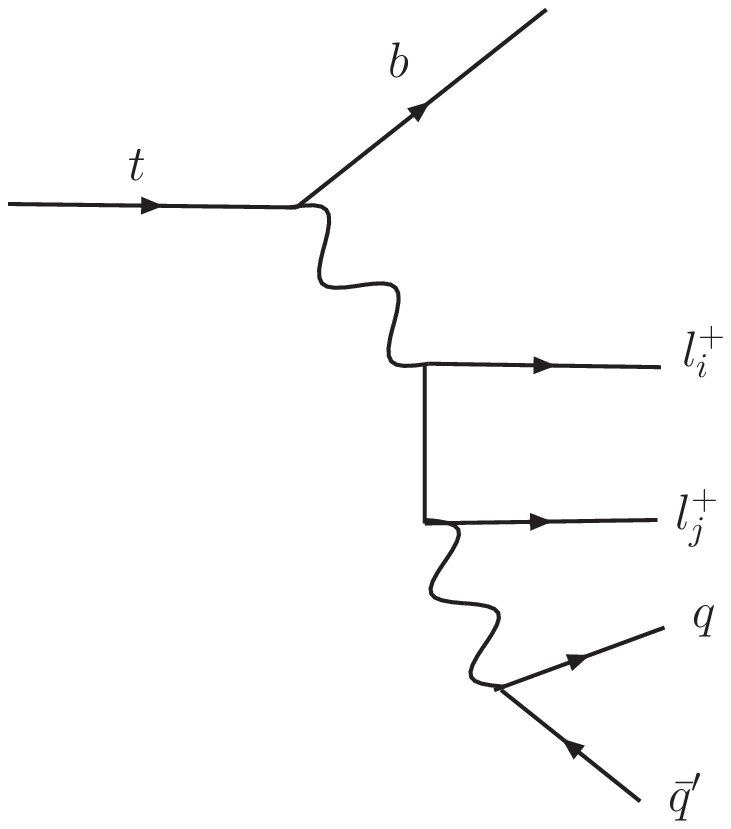}
\includegraphics[scale=1,width=5cm]{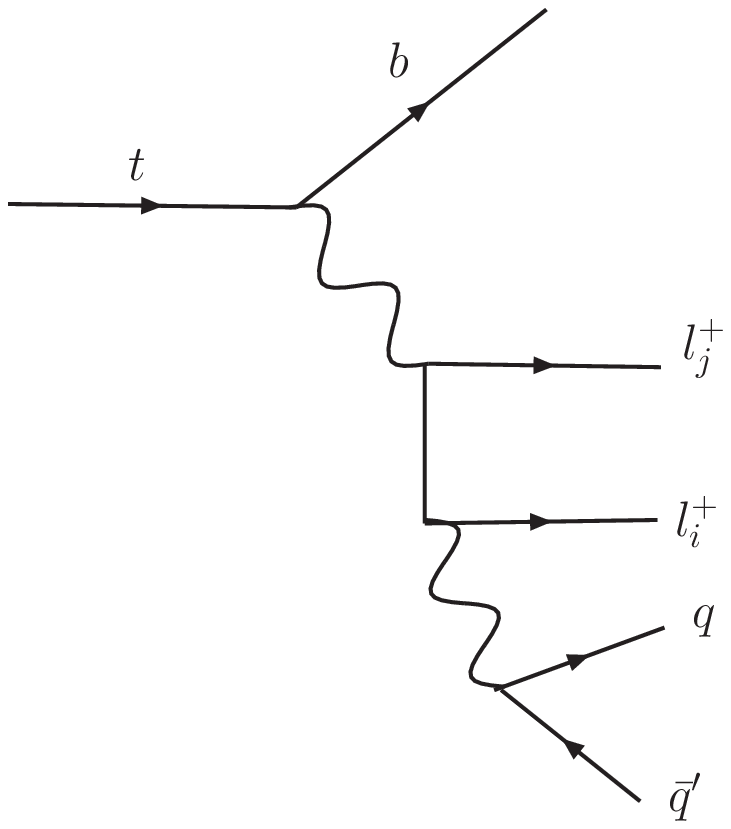}
\caption{Like-sign lepton pair production in top quark decay: 
$t\to b l^+ l^+ q \bar{q}'$ }
\label{tll}
\end{figure}

As discussed in the introduction, if a Majorana neutrino occurs as intermediate state in  $W$ decay, we will encounter a same-sign dilepton ($\Delta L=2$) final state as
$$W^\pm \rightarrow l^\pm N \rightarrow l^\pm l^\pm (W^\mp)^*.$$
To avoid combinatorial problem in lepton final states, we require the $W^*$ to decay hadronically. 
Therefore, for the top quark decay through a Majorana neutrino, we are interested in the cascade as (Fig. \ref{tll}) 

\be
\label{tdll}
t(p) \to b(p_b) +l_i^+(l_i) +l_j^+(l_j)+ q(j_1)+ \bar{q}'(j_2),
\ee
where $p$, $p_b$, etc. denote the 4-momentum of the corresponding particles. 
The differential decay width for this channel is given as:
\be
d\Gamma_{t\to bl^+l^+q\bar{q}'}=\frac{1}{2m_t} 
|{\cal M}_{t\to bl^+l^+ q\bar{q}'}|^2 d \text{PS}_5,
\ee
where $d\text{PS}_5$ denotes the 5-body phase space, and $m_t$
represents the top quark mass.

The corresponding matrix element squared is given as follows: 
\be 
\label{amtll}
|{\cal M}_{t\to bl^+l^+ q\bar{q}'}|^2 = 
\frac{g^8 N_c M_N^2 |V_{iN}V_{jN}|^2 |V_{tb}|^2 |V_{qq'}|^2 (1-\frac{1}{2} \delta_{ij}) 
}{
[(p_w^2-m_W^2)^2+\Gamma_W^2 m_W^2][(p_w'^2-m_W^2)^2+\Gamma_W^2 m_W^2]}
\Big\{
F- \frac{G}{D_{11} D_{22}} + \Big[{l_i \leftrightarrow l_j}\Big]
\Big\},
\ee
where $g=e/\sin \theta_W$, $N_c=3$, $\Gamma_{W}(m_W)$ is the width(mass)
of the $W$ boson, $M_N$ is the heavy neutrino mass, $V_{tb/qq'}$ is the 
CKM matrix elements and $V_{iN}$ is the rotation od neutrino mass eigen states defined in the Eq.~\ref{mix}.

Majorana neutrino $N$ width is 
\be
\Gamma_N=\sum_{i=e,\mu,\tau} 18 |V_{iN}|^2 \left({G^2_F M^5_N\over 192 \pi^3}\right) ,~~~~~(M_N < m_W)
\ee
and
\be
\Gamma_N=\sum_{i=e,\mu,\tau} |V_{iN}|^2 \left({G_F M^3_N\over 8}\right) ,~~~~~(M_N > m_Z, m_H).
\ee
Since the total width of Majorana neutrino contains a factor as $\sum_{i=e,\mu,\tau}|V_{iN}|^2$ and it will appear in the Majorana neutrino
propagator, we follow \cite{Han:2006ip} to define an effective mixing parameter as
\begin{equation}
S_{ij}={|V_{iN}V_{jN}|^2\over\sum_{i=e,\mu,\tau}|V_{iN}|^2}~.
\end{equation}
We then can then normalize the physics variables by $S_{ij}$. 
  
\begin{figure}[!tb]
\includegraphics[scale=1,width=8cm]{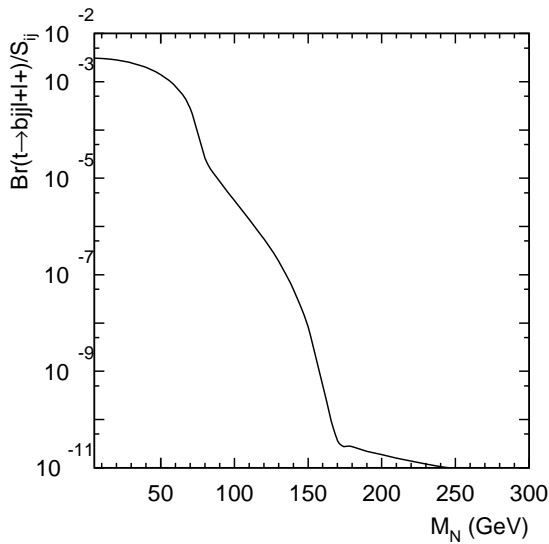}
\caption{Normalized Decay branching ratio of the process $t\to b l^+ l^+ q \bar{q}'$ } 
\label{tllwid}
\end{figure}

The normalized branching ratio for $t\rightarrow b l^+l^+ jj$ vs the Majorana neutrino mass $M_N$
is plotted in Fig. \ref{tllwid}. 
For $M_N$ below $m_W$, the on-shell decay of $W$ into Majorana neutrino can be as large as $0.02 S_{ij}$.
If the Majorana neutrino is within $m_W< M_N<m_t$,  the top three body decay $t\rightarrow bl^+ N$ with 
onshell Majorana neutrino varies between $10^{-5} S_{ij}$ and $10^{-10} S_{ij}$. 
If $M_N>m_t$, the decay BR is less $10^{-10} S_{ij}$ and irrelavant to our search. 
For the $M_N>m_t$, the results agree with those obtained in Ref.\cite{Bar-Shalom:2006bv}. However, in the
region where $M_N<m_W$, the results disagree with those in Ref.\cite{Bar-Shalom:2006bv}.

\section{Discovery at the LHC}

The LHC is a top rich environment, which enables us to use the
$t\bar{t}$ events to investigate the Majorana neutrino signals. 
From Fig. \ref{tllwid}, if $N$ is off-shell produced or from top three body decay, the chance to discover this channel will be extremely tiny. 
We focus on the region where $N$ can be on-shell produced from $W$. The most striking signature for the Majorana neutrino 
production is from a same sign dilepton $l^\pm l^\pm$. Therefore the visibility of two isolated same-sign leptons is essential 
to our search. If $M_N$ is in close degeneracy with $m_W$, the lepton from $W$ will be extremely soft and very hard to detect. 
At another extreme where the $N$ is very light, the decay products from the $N$ will be also soft and the $N$ boost will make the lepton and hadrons 
collimated and hence difficult to isolate. Therefore, the most optimistic region will be $M_N$ within 15 to 65 GeV range.  
We choose a $M_N = 15$ GeV for the purpose of illustration. 

The total cross section of $t\bar{t}$ production at hadron colliders is defined
as follows
\be
d\bar{\sigma} = \int d x_1dx_2 f_{a/A}(x_1) f_{b/B}(x_2) d\hat{\sigma}_{ab\to t\bar{t}},
\ee
where $f(x)$ denotes the parton distribution function, and $ d\hat{\sigma}$ represents for
the differential cross section at parton level. At Tevatron and LHC, there are 
two dominant partonic processes:
\bea
q(p_1)+\bar{q}(p_2)\to t(k_1)+\bar{t}(k_2),\\
g(p_1)+g(p_2) \to t(k_1) + \bar{t}(k_2)
\eea
at leading order of QCD. Their differential cross section 
are given as follows\cite{Nason:1987xz}:
\be
d\hat{\sigma}_{ab\to t\bar{t}}= \frac{1}{2\hat{s}} |{\cal M}_{ab}|^2 
d \text{PS}_2,
\ee
where $\hat{s}=(p_1+p_2)^2$, $d \text{PS}_2$ is the two body phase space, and
the corresponding matrix elements squared are as follows:
\bea
|{\cal M}_{q\bar{q}}|^2 &=& \frac{g_s^2 (N_c^2-1)}{4N_c^2} 
\Big\{ 2-\beta^2(1-y^2)\Big\}, \nonumber \\
|{\cal M}_{gg}|^2 &=& \frac{g_s^2\left[ N_c^2 
\left(1+\beta^2 y^2\right)-2\right]}{2N_c (N_c^2-1) (1-\beta^2 y^2)^2} 
\Big\{
1+2\beta^2(1-y^2)- \beta^4 \left[ 1+(1-y^2)^2\right]
\Big\},
\eea
with $y=\hat{p}_1\cdot \hat{k}_1$, and $\beta=\sqrt{1-4m_t^2/\hat{s}}$ and $N_c=3$.

To minimize the lepton combinatorial problem, we require the second top to decay hadronically. At leading order,
the final state consists of 6 jets (two of them are b-jets) and same-sign dilepton with no significant $\cancel{E}_T$, 
\begin{equation}
\label{A:hh2tt}
pp \to t\bar{t} \to b\bar{b} + l^\pm l^\pm  + j_1 j_2 j_3 j_4.
\end{equation}
$t\bar{t}$ production involves very active QCD radiation and the jets from virtual $W$ decay 
are as soft as the radiation jets. It is hard to require inclusive signature of exactly 6 jets.
Therefore, at the trigger level, we do not impose the 6 jets requirement and we use the two top
reconstruction to categorize jets. 

\begin{figure}[!tb]
\includegraphics[scale=1,width=8cm]{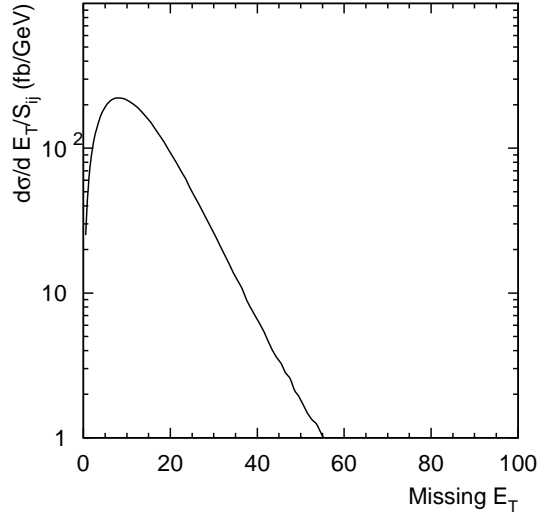}
\caption{$\cancel{E}_T$ distribution of $pp\rightarrow t\bar{t}\rightarrow 6j+l^\pm l^\pm$ after detector smearing effect, normalized by the mixing
parameter $S_{ij}$ }
\label{miss}
\end{figure}

The key feature for this channel is the same-sign dilepton with no missing energy associated. However, due to
the measurement of jet energy or electromagnetic energy of leptons, $\cancel{E}_T$ may appear. To simulate the detector 
effects on the energy-momentum measurements, we smear the electromagnetic energy
and the muon momentum by a Gaussian distribution whose width is parameterized as \cite{CMS}
\begin{eqnarray}
{ \Delta E\over E} &=& {a_{cal} \over \sqrt{E/{\rm GeV}} } \oplus b_{cal}, \quad
a_{cal}=5\%, b_{cal}=0.55\% ,
\label{ecal}\\
{\Delta p_T\over p_T} &=& {a_{track}p_T \over {\rm TeV}} \oplus {b_{track}\over \sqrt{\sin{\theta}} }, \quad
 a_{track}= 15\%,b_{track} =0.5\%.
\end{eqnarray}
The jet energies are also smeared using the same Gaussian formula as in Eq.~(\ref{ecal}),
but with  \cite{CMS}
\begin{equation}
a_{cal}=100\%,\quad  b_{cal}=5\%.
\end{equation}
The smearing simulation in Fig.\ref{miss} shows that $\cancel{E}_T$ cannot be neglected. 
We require that there is no significant $\cancel{E}_T$ as
\be
\cancel{E}_T < 25 \text{GeV}
\ee

\begin{figure}[!tb]
\includegraphics[scale=1,width=8cm]{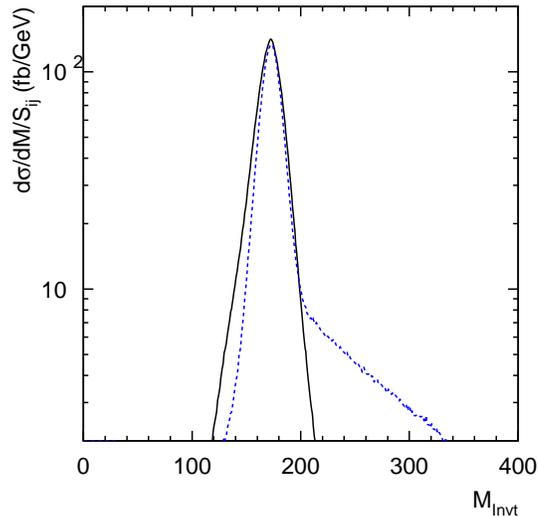}
\caption{Invariant mass distribution of two fully reconstructed tops normalized by the mixing parameter $S_{ij}$. Solid line
corresponds to the first-reconstructed hadronic top and dash line corresponds to the leptonic top.}
\label{2top}
\end{figure}

We propose the basic cuts as
\begin{itemize}
\item same-sign dilepton with $p_T(l) > 10~\text{GeV}$ and $|\eta(l)|< 2.8$
\item at least 3 jets with $p_T(j) > 50$ GeV and $|\eta(j)| < 3.0$
\item $\cancel{E}_T < 25$ GeV
\item $R_{jl}$, $R_{jj}$, $R_{ll}  > 0.4$
\end{itemize}
We only require 3 hard jets at the trigger level. However, to identify the signal, 
the first step is to reconstruct two tops. 
We demand two $b$-tagged jets, plus 4 more jets, along with the two same-sign dilepton.
By first taking the three-jet invariant mass which is closest to $m_t$, one can group
the three jets from hadronic top decay then group everything else together to 
construct invariant mass. Fig. \ref{2top} shows the simulated signal event following
this jet categorization procedure.  
The top reconstruction serves two purposes. One is to identify the event and remove the 
multijets+$W^\pm W^\pm$ or $t\bar{t}W^\pm$ background. By requiring the second invariant mass
\be
|M_{\rm inv}-m_t|< 30 \text{GeV}~,
\ee
one can argue that there is no standard model background and the signal is essentially event counting. 

The second purpose is to properly group the jets. In this channel, there is no significant missing 
$E_T$ in the final states. This provides us a way using only
invariant mass variables to fully reconstruct the events. 

\begin{figure}[!tb]
\includegraphics[scale=1,width=8cm]{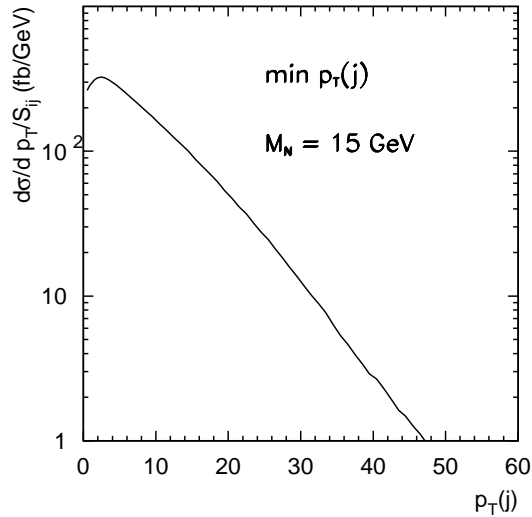} 
\caption{The minimal jet transverse momentum distribution ${\rm min}\{p_T(j)\}$ of $N\rightarrow l jj$ normalized by the mixing parameter $S_{ij}$. }
\label{ptj}
\end{figure}

\begin{figure}[!tb]
\includegraphics[scale=1,width=8cm]{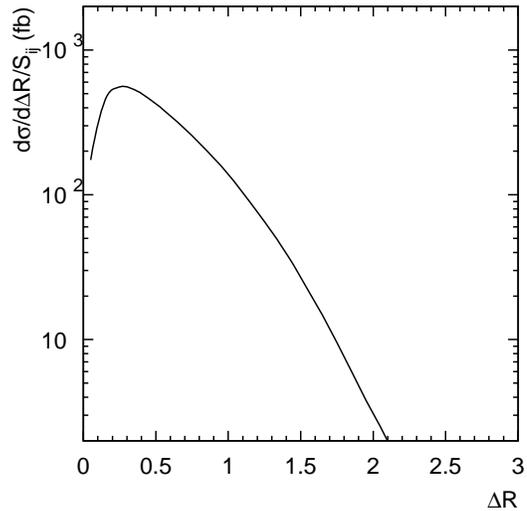}
\caption{The minimal separation between lepton and jet distribution ${\rm min}\{\Delta R_{lj}\}$ normalized by the mixing parameter $S_{ij}$.}
\label{dr}
\end{figure}

In the case of $M_N= 15$ GeV, the decay products from $N \rightarrow l jj$ will 
be very soft, and $W$'s from $t\rightarrow bW$ are on-shell produced. Then 
the $N$ boost will enhance the jet $p_T$ 
and make the $N \rightarrow l jj$ collimated in the $N$ boost direction. 
Fig. \ref{ptj} shows the ${\rm min}\{p_T(j)\}$ in the event and Fig. \ref{dr}
shows the ${\rm min}\{\Delta R_{lj}\}$ due to $N$ boost.
We define a cone of all these soft jets and one lepton then construct the invariant 
mass, which gives us the $M_N$.
\begin{figure}[!tb]
\includegraphics[scale=1,width=8cm]{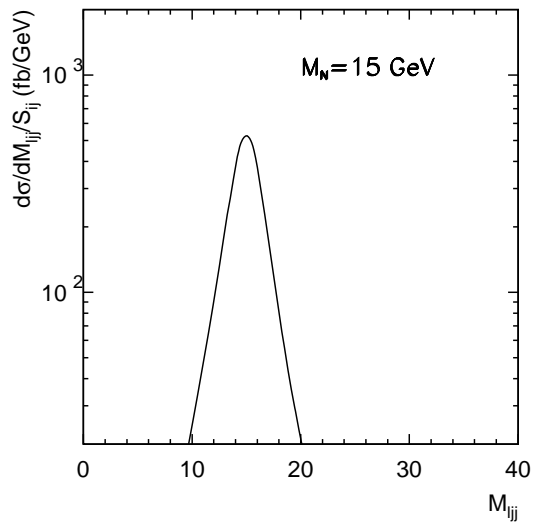}
\caption{Invariant mass distribution of of jets and lepton that reconstruct $M_N$ normalized by the mixing parameter $S_{ij}$.}
\label{fig:mn}
\end{figure}

\begin{figure}[!tb]
\includegraphics[scale=1,width=8cm]{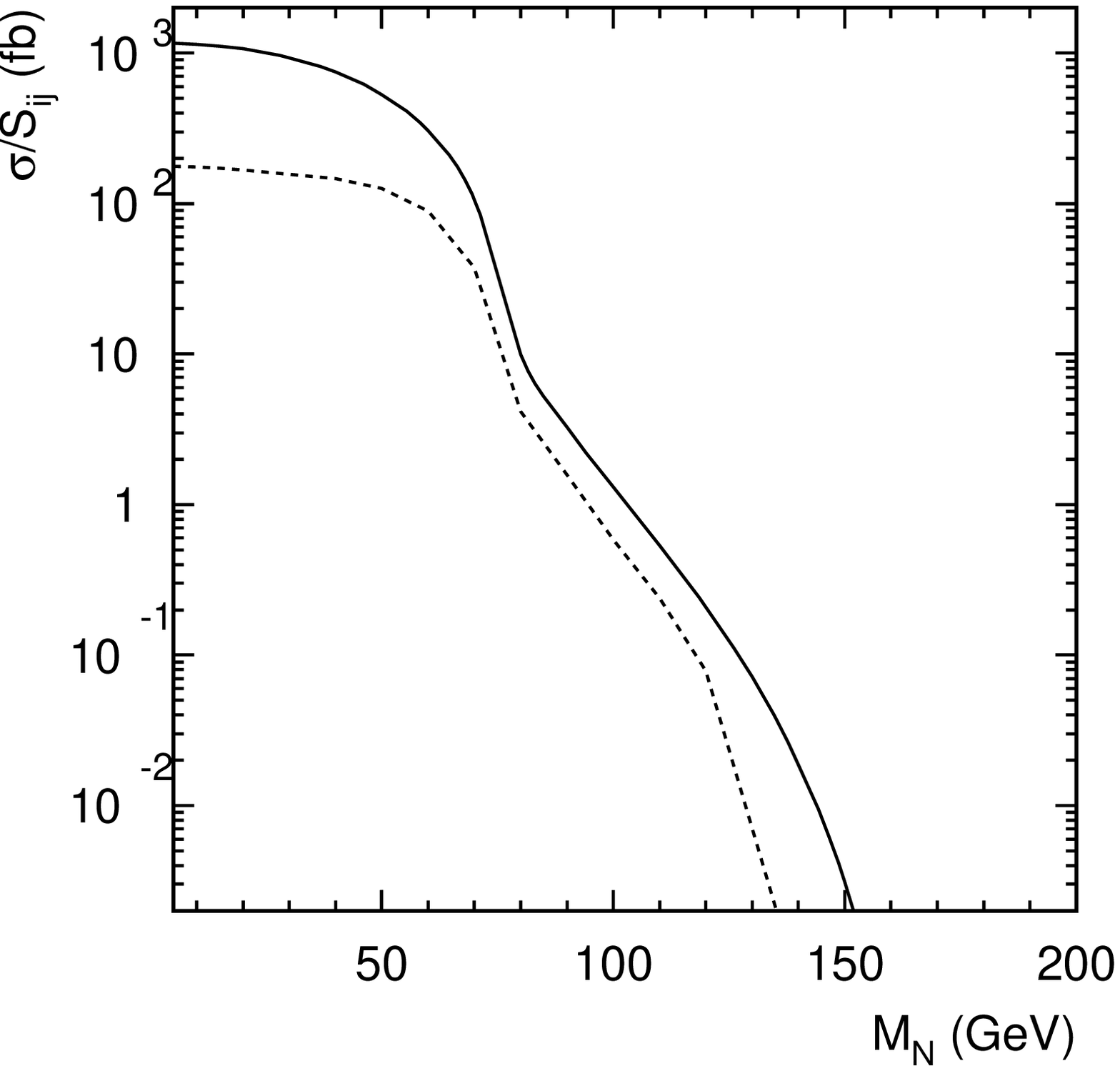}
\includegraphics[scale=1,width=8cm]{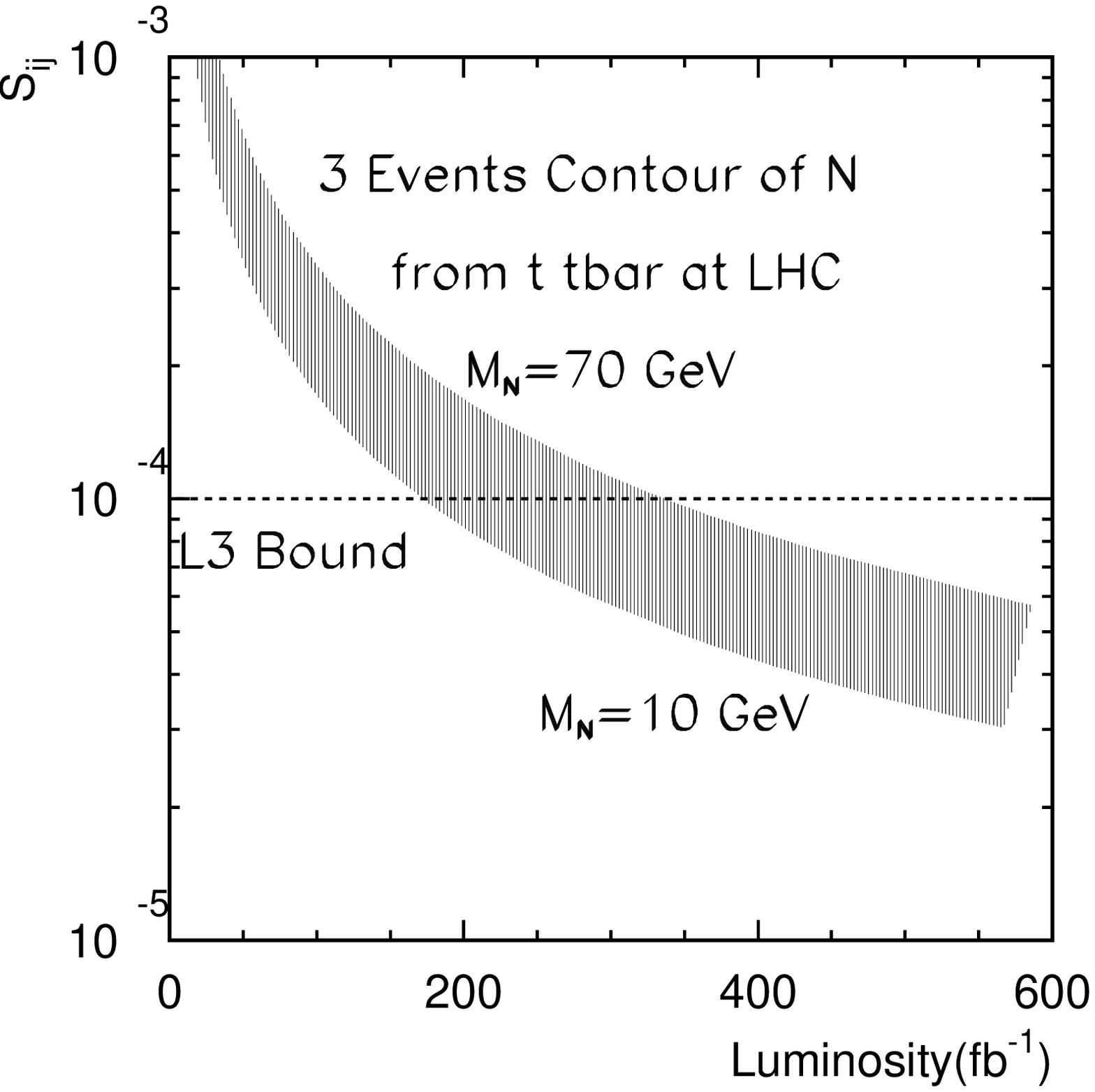}
\caption{Left: Cross section of $pp\to t\bar{t}\to b\bar{b} ll jj$ at LHC. 
Solid/dashed line without/with cuts, normalized by the mixing parameter $S_{ij}$. Right: 3 events contour of $N$ decay from $t\bar{t}$ pair at the LHC}
\label{figtot}
\end{figure}

To illustrate other mass region, 
we show in Fig.~\ref{figtot} the total cross section of the top quark decay to a Majorana
neutrino versus $M_N$ at the LHC energy. The solid (dashed) curve represents the
cross section without (or with) the basic kinematic cuts as
\begin{itemize}
\item same-sign dilepton with $p_T(l) > 10~\text{GeV}$ and $|\eta(l)|< 2.8$
\item 6 jets with $p_T(j) > 15$ GeV and $|\eta(j)| < 3.0$
\item $\cancel{E}_T< 25$ GeV
\item $R_{jl}$, $R_{jj}$, $R_{ll}> 0.4$
\end{itemize}

As we argued earlier, the two top reconstruction requirement reduces the SM background to a negilible level so the signal is just 
event-counting. To summarize the reach of different mass of $N$, in Fig.~\ref{figtot}, we also show the 3 events contour of this channel at the LHC. 
Since $M_N$ is fully reconstructable, one can use the event-counting to probe the effective mixing parameter $S_{ij}$.

\section{Summary}
Due to the large event sample size of the top quarks at the LHC, we consider the signal of a Majorana neutrino
from top-quark decay. The signature is same-sign dilepton plus multi-jets with no significant missing energy.
The most optimistic region lies where the Majorana neutrino mass is between $15-65$ GeV. For 300 fb$^{-1}$ integrated luminosity,
it is possible to probe the effective mixing parameter $S_{ij}$  to ${\mathcal O}(10^{-5})$. Since the bounds on $|V_{eN}|^2$ already ruled out the reach at LHC, if one can
identify $e^\pm e^\pm$ final states in the top decay chain, it will be from $\tau^\pm \tau^\pm$ leptonic decay. In order to get
a better sensitivity than the LEP experiments on  $|V_{\mu N}|^2$,$|V_{\tau N}|^2$, it will require total
integrated luminosity to be higher than 200 fb$^{-1}$.

\begin{acknowledgments}
We would like to thank Tao Han for useful discussion and early collaboration of this work. We are grateful for Bin Zhang 
who has provided code for computing Majorana neutrino width. We would also thank Tao Han and William Klemm for careful reading the manuscript and providing useful suggestions. Si is supported by NSFC, NCET and Huoyingdong Foundation and in part by the Project of Knowledge Innovation Program (PKIP) of Chinese Academy of Sciences. Si would also like to acknowledge the hospitality of the Phenomenology Institute, University of Wisconsin-Madison while the work was initiated. KW was supported in part by the U.S.~Department of Energy under grants DE-FG02-95ER40896 and the Wisconsin Alumni Research Foundation. KW is also supported by the World Premier International Research Center Initiative (WPI Initiative), MEXT, Japan. 
\end{acknowledgments}

\begin{appendix}

\section{Top decay to a Majorana neutrino}

In this appendix, we give the derivation of the Majorana  neutrino decay partial width calculation.
As is well known, the top quark width($\Gamma_t$) is much smaller
than its mass($m_t$). The Leading Pole Approximation(LPA) can then be applied.
Under LPA, the cross section of the process (\ref{A:hh2tt}) can be factorized
into two parts: $t\bar{t}$ pair production and top quark decays, i.e.,
\begin{eqnarray}
d\sigma &=& \frac{1}{\Gamma_t^2} d\bar{\sigma}_{pp/p\bar{p}\to t\bar{t}}
\Big\{
d\Gamma_{t\to b l^+ l^+ j_1 j_2} d\Gamma_{\bar{t}\to \bar{b} j_3 j_4}
+ d\Gamma_{\bar{t}\to \bar{b} l^- l^- j_1 j_2} d\Gamma_{t\to b j_3 j_4}
\Big\},
\end{eqnarray}
where $d\bar{\sigma}$ denotes the differential cross section for $t\bar{t}$ production,
and $d\Gamma$ is the corresponding top quark decay differential decay width. $\Gamma_t$ is
total decay width of top quark.

\be
\label{A:tdll}
t(p) \to b(p_b) +l_i^+(l_i) +l_j^+(l_j)+ q(j_1)+ \bar{q}'(j_2),
\ee
where $p$, $p_b$, etc. denotes the 4-momentum of the corresponding particles. Its 
differential decay width is given as follows:
\be
d\Gamma_{t\to bl^+l^+q\bar{q}'}=\frac{1}{2m_t} 
|{\cal M}_{t\to bl^+l^+ q\bar{q}'}|^2 d\text{PS}_5,
\ee
where $d\text{PS}_5$ denotes the 5-body phase space. The quark pair
$q\bar{q}'$ is mainly $d\bar{u}$ and $s\bar{c}$. The corresponding matrix element squared is given as follows:
\be 
\label{A:amtll}
|{\cal M}_{t\to bl^+l^+ q\bar{q}'}|^2 = 
\frac{g^8 N_c M_N^2 |V_{iN}V_{iN}|^2|V_{tb}|^2 |V_{qq'}|^2 (1-\frac{1}{2} \delta_{ij}) 
}{
[(p_w^2-m_W^2)^2+\Gamma_W^2 m_W^2][(p_w'^2-m_W^2)^2+\Gamma_W^2 m_W^2]}
\Big\{
F- \frac{G}{D_{11} D_{22}} + \Big[{l_i \leftrightarrow l_j}\Big]
\Big\},
\ee
where $g=e/\sin \theta_W$, and
\bea
F& =& \frac{2 (l_j j_1)}{D_{11}} \Big\{
4 (p j_2 )(p_b l_i) - \frac{2m_t^2}{m_W^2} \Big[
(j_2  p_b)[ l_i\cdot (p-2p_b)] + (p j_2)(p_b l_i)-(j_2 l_i)(p p_b) \Big]
\nonumber \\
&& ~~~~~~~~~ + \frac{m_t^2 (p p_b)}{m_W^4}\Big[
2(j_2 p_w)( l_i p_w) - (l_i j_2)[m_t^2- 2 (p p_b)] \Big]
\Big\}, \nonumber \\
D_{11}&=& D_1^2 + \Gamma_N^2M_N^2, ~~~~~D_1 = (l_i-p_w)^2-m_W^2, ~~~
p_w=p-p_b, \nonumber \\
D_{22}&=&D_2^2  + \Gamma_N^2M_N^2, ~~~~~D_2 = (l_j-p_w)^2-m_W^2, ~~~
p_w'=j_1+j_2.
\eea
We use the notation $(p p_b) \equiv p\cdot p_b$, etc. 
The term $G$ in Eq.~(\ref{amtll}) is from the interference between the two diagrams of fig.\ref{tll}:
\be
G  =\Big[ D_1 D_2+ \Gamma_N^2  M_N^2\Big] G_1 +\Big[D_1- D_2 \Big]\Gamma_N M_N  G_2,
\ee
where
\bea
G_1 &=& 4 (p j_2) \Big\{ (l_j j_1)(p_b l_i) +
(j_1 l_i)(p_b l_j) -(p_b j_1)(l_i l_j)
\Big\} \nonumber \\
&+&\frac{ m_t^2 (p p_b)}{m_W^4} \Big\{
2 (l_i j_1)\Big[2 (j_2 p_w) (l_j p_w) - (l_j j_2)[m_t^2-2 (p p_b)] \Big] \nonumber \\
&-&(l_i l_j) \Big[ 2(j_1 p_w) (j_2 p_w)-(j_1 j_2) [m_t^2-2 (p p_b)]\Big]\Big\} \nonumber \\
&+&  \frac{2m_t^2}{m_W^2} \Big\{
 - 2 (j_1 l_j)\Big[ (p_b j_2) [l_i\cdot(p-2p_b)]+ (p j_2)(p_b l_i)-  (j_2 l_i) (p p_b)\Big] \nonumber \\
&+&  (l_i l_j)\Big[ (p_b j_1) [j_2\cdot (p-2 p_b)] +(p j_1)(j_2 p_b) -  (j_1 j_2)(p p_b)\Big] \Big\} \nonumber \\
G_2&=&(l_i l_j) \Big\{  \omega \epsilon_{j_1 j_2 (l_i-l_j) p_b}
-\frac{2m_t^2 (p p_b)}{m_W^4} \epsilon_{j_1 j_2 l_i l_j} \Big\}
-2 (j_1 l_i) \Big\{\omega \epsilon_{(j_1-l_i) j_2 l_j p_b} 
+\frac{2m_t^2 (p p_b)}{m_W^4} \epsilon_{j_1 j_2 l_i l_j}\Big\} \nonumber \\
&&+ \epsilon_{j_1j_2l_i l_j} \Big\{ 2 \omega  (j_2 p_b) + (1+\frac{m_t^2}{m_W^2}) (p p_b)\Big\}
+ 2 (3-\frac{m_t^2}{m_W^2}) (j_2 l_i) \epsilon_{j_1(l_i+j_2) l_j p_b}
+4 (p_b l_i) \epsilon_{j_1j_2l_jp_b} \nonumber \\
&&+(3-\frac{m_t^2}{m_W^2} )(j_1 j_2) \epsilon_{(j_1+j_2)l_i l_j p_b}
+ 2 \Big\{ (j_1 p_2) \epsilon_{j_2 l_i l_j p_b}+(j_2 p_b) \epsilon_{j_1 l_i l_j p_b}\Big\},
\eea
\be
\omega =\frac{m_t^2-m_W^2}{m_W^2}, ~~~~~~~~~
\epsilon_{j_1 j_2 l_i l_j} \equiv \epsilon_{\mu\nu\rho\sigma} j_1^{\mu} j_2^{\nu} l_i^{\rho} l_j^{\sigma}.
\ee

\end{appendix}

\end{document}